\documentclass[twocolumn]{aastex631}

\usepackage{graphicx}	
\usepackage{amsmath}	
\usepackage{hyperref}
\usepackage{cprotect}
\usepackage{multirow}

\newcommand{\jr}{\color{black}}

\begin{document}

\title{Most Super-Earths Have Less Than 3\% Water}

\author[0000-0001-7615-6798]{James G. Rogers}
\affiliation{Institute of Astronomy, University of Cambridge, Madingley Road, Cambridge CB3 0HA, United Kingdom}

\author[0000-0001-6110-4610]{Caroline Dorn}
\affiliation{Institute for Particle Physics and Astrophysics, ETH Zurich Department of Physics, Wolfgang-Pauli-Strasse 27, CH-8093 Zurich, Switzerland}

\author[0009-0004-7590-3818]{Vivasvaan Aditya Raj}
\affiliation{Department of Earth, Planetary, and Space Sciences, The University of California, Los Angeles, 595 Charles E. Young Drive East, Los Angeles, CA 90095, USA}

\author[0000-0002-0298-8089]{Hilke E. Schlichting}
\affiliation{Department of Earth, Planetary, and Space Sciences, The University of California, Los Angeles, 595 Charles E. Young Drive East, Los Angeles, CA 90095, USA}

\author[0000-0002-1299-0801]{Edward D. Young}
\affiliation{Department of Earth, Planetary, and Space Sciences, The University of California, Los Angeles, 595 Charles E. Young Drive East, Los Angeles, CA 90095, USA}



\begin{abstract}
Super-Earths are highly irradiated, small planets with bulk densities approximately consistent with Earth. We construct combined interior-atmosphere models of super-Earths that trace the partitioning of water throughout a planet, including an iron-rich core, silicate-rich mantle, and steam atmosphere. We compare these models with exoplanet observations to infer a $1\sigma$ upper limit on total water mass fraction of $\lesssim 3\%$ at the population level. We consider end-member scenarios that may change this value, including the efficiency of mantle outgassing, escape of high mean-molecular weight atmospheres, and increased iron core mass fractions. Although our constraints are agnostic as to the origin of water, we show that our upper limits are consistent with its production via chemical reactions of primordial hydrogen-dominated atmospheres with magma oceans. This mechanism has also been hypothesised to explain Earth's water content, possibly pointing to a unified channel for the origins of water on small terrestrial planets.
\end{abstract}



\section{Introduction} \label{sec:intro}
Super-Earths offer a unique insight into planet formation. With orbital periods $\lesssim 100$~days and sizes $\lesssim 1.8 \text{R}_\oplus$, super-Earths are among the most common class of planet in our galaxy \citep[e.g.][]{Howard2012,Fressin2013,Petigura2013,Mulders2015,Zink2023}. Precise mass and radius measurements have revealed their bulk densities to be ``Earth-like'' \citep[e.g.][]{Wu_Lithwick2013,Weiss2014,Rogers2015,Wolfgang2016,Chen2017}, prompting questions as to whether Earth's interior composition is a natural outcome of planet formation.

Perhaps the most convenient definition of super-Earths is that they sit below the radius valley. The radius valley is an observed paucity in planet occurrence $\sim 1.8 \text{ R}_\oplus$, although its exact location varies with orbital period and stellar mass \citep[e.g.][]{Fulton2017,Petigura2018,VanEylen2018,Petigura2022,Ho2023}. Above the radius valley sits the population of sub-Neptunes, whose bulk density is markedly reduced when compared to super-Earths. Inferring the composition of sub-Neptunes is fraught with degeneracy \citep[e.g.][]{Rogers2015,Bean2021,Rogers2023b}. This arises since there are expected to be $4$ major chemical reservoirs within a sub-Neptune: iron, silicates, H$_2$O-dominated volatiles, and hydrogen. Whilst it is known that a significant hydrogen mass fraction is required to reproduce the densities of the largest sub-Neptunes \citep[e.g.][]{JontofHutter2016}, attempting to constrain the mass fractions of each reservoir when only utilising $2$ observables, being planetary mass and size, is an under-constrained mathematical problem. 

To alleviate some of the degeneracy of sub-Neptune composition, one can utilise planet structure and evolution models. However, different models currently predict different interior compositions. On the one hand, models of atmospheric evolution including escape, suggest that super-Earths and sub-Neptunes formed from the same underlying planet population of hydrogen-dominated atmospheres atop Earth-like interiors \citep[e.g.][]{Owen2013,LopezFortney2013,Ginzburg2018,Gupta2019,Rogers2021}. On the other hand, planet population-synthesis models predict a prominent population of water worlds, defined as planets with a water mass fraction of several tens of per cent \citep[e.g.][]{Mordasini2009,Raymond2018,Zeng2019,Venturini2016,Mousis2020,Burn2024}, arising due to formation beyond the water-ice line, followed by large-scale migration \citep[e.g.][]{Lodders2003,Bitsch2015,Venturini2020,Bruger2020}. 

Perhaps the most promising avenue for breaking compositional degeneracy is with models of planet structure \textit{and} chemical composition. For example, the requirement of hydrostatic and chemical equilibrium provides more stringent constraints on the possible outcomes of planet formation \citep[e.g.][]{Schlichting2022,Young2023,Rogers2024b}. Future atmospheric characterisation of sub-Neptune atmospheres may inform these models \citep[see][]{Benneke2024}, although for now, our knowledge of sub-Neptune compositions remains uncertain.

Super-Earths, on the other hand, provide a simpler chemical system to solve. What is clear for planets existing below the radius valley is that it is challenging to explain the existence of a significant, long-lived hydrogen-dominated atmosphere. Their notably higher bulk densities provide observational evidence that they do not host large hydrogen atmospheres, which is reinforced by theoretical insights into atmospheric escape under intense stellar irradiation \citep[e.g.][]{OwenJackson2012,Kubyshkina2018,Schulik2022,Caldiroli2022}. In short, planets below the radius valley are typically smaller in mass and experience higher equilibrium temperatures, meaning that they are very unlikely to host a hydrogen-dominated atmosphere if indeed they ever had one \citep[see also][]{Lee2022,Rogers2024a}. 

From the original list of $4$ chemical reservoirs for sub-Neptunes, the lack of atmospheric hydrogen for super-Earths reduces the list to $3$ major chemical components: iron, silicates and H$_2$O-dominated volatiles. For observed mass and radius, this still yields an under-constrained problem. However, it allows one to explore parameter space in a more straightforward manner.

In this study, we focus on the water content of super-Earths. Water is evidently of major interest for habitability, however, it is also important in the context of understanding planet formation and evolution. Formation scenarios resulting in several tens of per cent in water mass fraction clearly require delivery of water from ice-rich pebbles or planetesimals formed beyond the ice line \citep[e.g.][]{Mordasini2009,Raymond2018}. However, recent studies have shown that water can be \textit{endogenously} produced as a result of the chemical equilibration of hydrogen-dominated atmospheres and magma oceans \citep[e.g.][]{Kite2021,Schlichting2022}. The total water mass fraction produced through this mechanism is typically of order a few \% \citep{Schlichting2022,Rogers2024b}. A similar debate exists for the origins of water on Earth, either delivered \citep[e.g.][]{Raymond2004,Raymond2009} or chemically produced \citep[e.g.][]{Wu2018,Young2023}. 


In this study, we constrain the water content of super-Earths by constructing simple physical and chemical models that account for the partitioning of water between a planet's metal core, silicate mantle and atmosphere. We compare these models with observed planetary masses and radii in order to place upper limits on water mass fractions at the population level. Crucially, we are agnostic as to the origin of the water i.e. whether it was delivered or endogenously produced. We also explore processes that may change our statistical constraints, including atmospheric escape of high mean-molecular weight atmospheres and mantle outgassing. For a given water mass fraction, these scenarios can change the bulk density of a super-Earth and hence must be included when performing this analysis.

We simplify the problem by choosing to model planets with an Earth-like iron mass fraction, which is supported by various lines of evidence, including \citet{Adibekyan2021a}, who showed that the majority of super-Earths have iron mass fractions consistent with, or less than that of Earth.  Other lines of evidence in favour of Earth-like interior compositions come from spectroscopic observations of polluted white dwarfs \citep[e.g.][]{Doyle2020,Trierweiler2023,LauraRogers2024} and broadband transit photometry of catastrophically evaporating rocky planets \citep{CamposEstrada2024}. In this initial study, we also choose to construct a simple chemical system which ignores carbon, nitrogen and sulphur-bearing species since these are unlikely to contribute a significant fraction of a planet's overall mass budget.

Our paper is structured as follows: in Section \ref{sec:Method}, we present a model for super-Earth interiors (\ref{method:interior}) and atmospheres (\ref{method:atmosphere}). We discuss our end-member scenarios, including the effects of atmospheric escape and mantle outgassing in Section \ref{sec:Scenarios}, with results presented in Section \ref{sec:ModelResults}. In Section \ref{sec:water_content}, we apply these models to infer upper limits on water mass fractions of the observed super-Earth population. We discuss data selection (\ref{sec:DataSelection}) and our statistical methodology (\ref{sec:ConstraintsMethod}), with results presented in Section \ref{sec:Results_data}. Discussion and conclusions are included in Sections \ref{sec:Discussion} and \ref{sec:Conclusions}, respectively.

\vspace{1cm}
\section{A model for super-Earths}\label{sec:Method}

\subsection{Interior Model} \label{method:interior}

We employ an interior model from \citet{Luo2024}, which is based on earlier versions of \citet{Dorn2021} and \citet{dorn_can_2015}. For this application, we focus on rocky worlds with varying amounts of water content, parameterised by a water mass fraction, $X_{\text{H}_2\text{O}} \equiv M_{\text{H}_2\text{O}} / M_\text{p}$, where $M_{\text{H}_2\text{O}}$ is the total mass of water and $M_\text{p}$ is the total planet mass. The water can be present in the metal core, the mantle, and the surface, depending on the specific thermal state of the planets. The main addition to the model compared to \citet{Luo2024} is the ability of the model to outgas water from the mantle. We describe more details in the following. 

We consider a core made of Fe, H, and O. For solid Fe, we use the equations of state for hexagonal close packed (hcp) iron \citep{hakim_new_2018,miozzi_new_2020}. For liquid iron with, and without H$_2$O, we use \citet{Luo2024}. The core thermal profile is assumed to be adiabatic throughout the core. At the core-mantle boundary (CMB), there is a temperature jump as the core can be hotter than the mantle due to the residual heat released during core formation. We follow \citet{stixrude_melting_2014} and add a temperature jump at the CMB temperature such that the temperature at the top of the core is at least as high as the melting temperature of the silicates.

The mantle is assumed to be made up of three major constituents, i.e., MgO, SiO$_2$, FeO. For the solid mantle, we use the thermodynamical model \verb|Perple_X| \citet{connolly_perplex}, to compute stable mineralogy and density for a given composition, pressure, and temperature, employing the database of \citet{stixrude_thermal_2022}. For pressures higher than $\sim125$~GPa, we define stable minerals \textit{a priori} and use their respective equation of states from various sources \citep{fischer_equation_2011, faik_equation_2018, hemley_constraints_1992, musella_physical_2019}. For the liquid mantle, we calculate its density assuming an ideal mixture of main components (Mg$_2$SiO$_4$,SiO$_2$,FeO) \citep{stewart_shock_2020,faik_equation_2018,melosh_hydrocode_2007,ichikawa_ab_2020} and add them using the additive volume law. Note that we use Mg$_2$SiO$_4$ instead of MgO since the data for forsterite has been recently updated for the high-pressure temperature regime, which is not available for MgO to our knowledge. The mantle is assumed to be fully adiabatic.

Water can be added to the mantle melts, while the solid mantle is assumed to be dry. The addition of water reduces the density, for which we follow \citet{bajgain_structure_2015} and decrease the melt density per wt\% water by $0.036$ g cm$^{-3}$. For small water mass fractions, this reduction is nearly independent of pressure and temperature.
The melting curve of mantle material is calculated for dry and pure MgSiO$_3$ to which the addition of water \citep{katz_new_2003} and iron \citep{Dorn2018} can lower the melting temperatures. Water within the core will also lower its melting temperature, for which we follow \citet{luo_majority_2024}. This water can also be in both liquid and solid phases.
The partitioning between mantle melts and the water layer is determined by a modified Henry’s law, for which we use the fitted solubility function of \citet{dorn_hidden_2021}. For the partitioning of water between iron and silicates, we follow \citet{luo_majority_2024}. For the equilibration pressure of water to partition between iron and silicates, we use half of the core-mantle boundary pressure, which is within typically discussed values for Earth ($0.3-0.6$). Varying this pressure would introduce overall small changes in the distribution of water.

Beyond the mantle, water can be in solid, supercritical, and gas phases. Whenever the water is in the gas (steam) phase, we use the description in Section \ref{method:atmosphere} to calculate its extent and evolution. We also consider solid and supercritical water phases, which are calculated using the EOS compilation in \citet{haldemann_aqua_2020}. 


\subsection{Atmospheric Evolution Model} \label{method:atmosphere}
The models of \citet{Luo2024} prescribe the partitioning of water between the interior and atmosphere of a rocky super-Earth. The presence of water within the interior increases its radius due to its reduced density when compared to silicate rock and iron. Although most of the water is sequestered in the interior, the presence of a small steam atmosphere also increases the radius of a planet. However, it is imperative to consider the thermodynamic evolution of atmospheres since they will radiatively cool and contract, thus changing in size as a function of time. Furthermore, this evolution will depend on atmospheric composition due to varying opacities and mean molecular weights. To account for this, we build semi-analytic radiative-convective models of steam atmospheres based on the equivalent case of hydrogen-dominated atmospheres used in \citet{Owen2017,Gupta2019,Rogers2021,Misener2022,Rogers2023b}. For the deep, convective interior, we assume an adiabatic density profile of the form:
\begin{equation} \label{eq:temperature_profile_adiabat}
    \rho(r) = \rho_\text{rcb}  \bigg[ 1 + \nabla_\text{ad} \frac{G M_\text{p}}{c_\text{s}^2 R_\text{rcb}} \bigg( \frac{R_\text{rcb}}{r} - 1 \bigg) \bigg]^{\frac{1}{\gamma - 1}},
\end{equation}
where the subscript `rcb' refers to quantities evaluated at the radiative-convective boundary. Here,  $\rho$ is the density, $\gamma$ is the adiabatic index, $\nabla_\text{ad} \equiv (\gamma - 1) / \gamma$ is the adiabatic temperature gradient, $M_\text{p}$ is the planet's mass and $R_\text{rcb}$ is the position of the radiative-convective boundary. Above the convective region sits a radiative layer, assumed to be isothermal at an equilibrium temperature of $T_\text{eq}$. The sound speed in Equation \ref{eq:temperature_profile_adiabat} is evaluated at the radiative-convective boundary, $c_\text{s}^2 \equiv k_\text{B} T_\text{eq} / \mu m_\text{H}$, where $m_\text{H}$ is the mass of a hydrogen atom. For a steam-dominated atmosphere, we adopt a mean-molecular weight of $\mu=18$, and an adiabatic index of $\gamma = 1.33$ \citep[see Appendix 1B of][]{Kempton2023}. Since most super-Earths exist at high equilibrium temperatures \citep[particularly once accounting for survey biases][]{Rogers2021}, we evaluate models with $T_\text{eq} > 500$~K such that water is very unlikely to condense out of the gas phase within an atmosphere.

At each timestep, we cool the model with a radiative luminosity given by:
\begin{equation}
    L_\text{rad} = \frac{64 \pi \nabla_\text{ad} G M_\text{p}  \sigma T_\text{eq}^4}{3 \, c_\text{s}^2 \, \kappa_\text{rcb} \, \rho_\text{rcb}},
\end{equation}
where $\kappa_\text{rcb}$ is the gas opacity evaluated at the radiative-convective boundary. Here we use the tabulated H$_2$O Rosseland mean opacities from \citet{Kempton2023}, which include Rayleigh scattering and molecular absorption. Once the new total energy of the system has been calculated, including thermal contributions from the rocky interior \citep[see detailed description in][]{Misener2022}, we solve for the new radiative-convective boundary radius $R_\text{rcb}$ and density $\rho_\text{rcb}$ for a given total planet mass and energy using \verb|scipy.fsolve| \citep{SciPy2020}. From there, we calculate the photospheric radius by finding the location within the radiative isothermal layer for which the optical depth to incoming radiation is $\tau=2/3$. We evolve the system with an adaptive-step predictor-corrector ordinary differential equation solver. Note that we do not incorporate atmospheric escape in the evolution models at this stage, but instead consider end-member scenarios based on the retention of atmospheres, as discussed in Section \ref{sec:Scenarios}.

\begin{figure}
	\includegraphics[width=1.0\columnwidth]{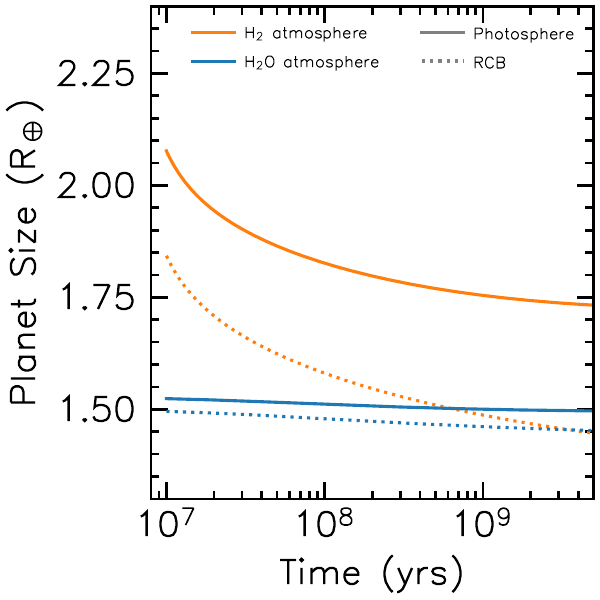}
    \centering
        \cprotect\caption{The atmospheric evolution for hydrogen and steam-dominated atmospheres. Each planet's photospheric and radiative-convective boundary (RCB) radii are shown in solid and dotted lines, respectively. An H$_2$-dominated atmosphere is shown in orange, and an H$_2$O-dominated atmosphere is shown in blue. Both planets are otherwise identical, with atmospheric mass fractions of $0.1\%$, initial atmosphere-base temperatures of $4000$~K, interior masses of $4M_\oplus$ and equilibrium temperatures of $1000$~K.} \label{fig:H2_vs_H2O_evolution} 
\end{figure}

For each model from Section \ref{method:interior}, we evolve its atmosphere for $5$~Gyrs, being the typical age of observed super-Earths \citep[e.g.][]{Berger2020}. We stress that the final size of planets after several Gyrs of evolution is insensitive to the exact choice in planet age due to the convergent nature of thermodynamic cooling and contraction \citep[e.g.][]{Rogers2023b}. For initial conditions, we set the base of the atmosphere to be at $4000$~K. Again, we highlight that the final size of planet atmospheres after Gyrs of evolution is insensitive to this value. {\jr{We confirm \textit{a posteriori} that these assumptions have a negligible effect on our conclusions by comparing calculations assuming different stellar ages of $1$~Gyr and $10$~Gyrs, which yield a difference in inferred water mass fraction $<0.3$~\% (see Section \ref{sec:ConstraintsMethod}). These simplifying assumptions would not hold when considering young planets, particularly for stellar ages $\lesssim 100$~Myrs when atmospheric contraction can be significant. Since typical ages of \textit{Kepler} and \textit{TESS} planet hosts are $>1$~Gyr, we can safely ignore this consideration.}} 

Figure \ref{fig:H2_vs_H2O_evolution} shows two example models of atmospheric evolution, highlighting the differences between H$_2$O steam and H$_2$ atmospheres in blue and orange, respectively. Both models are otherwise identical, with atmospheric mass fractions of $0.1$\%, interior masses of $4 \,\text{M}_\oplus$, interior radii $1.44 \,\text{R}_\oplus$, with initial atmospheric base temperatures of $4000$~K, orbiting Solar-mass stars with equilibrium temperatures of $1000$~K. For the H$_2$ atmosphere in Figure \ref{fig:H2_vs_H2O_evolution}, we adopt a mean molecular weight of $\mu = 2.35$ \citep{Anders1989} and a Rosseland mean opacity scaling of:
\begin{equation}
    \kappa = 1.29 \times 10^{-2} \bigg ( \frac{P}{1 \text{ bar}}\bigg)^\alpha \; \bigg ( \frac{T}{1000 \text{ K}}\bigg)^\beta \text{ cm}^2 \text{ g}^{-1},
\end{equation}
where $\alpha=0.68$ and $\beta=0.45$ \citep{RogersSeager2010,Owen2017}. Figure \ref{fig:H2_vs_H2O_evolution} shows each planet's size, including their photospheric and radiative-convective boundary radii as solid and dotted lines, respectively. The impact of mean-molecular weight is clear, with the H$_2$ model being significantly more inflated than the H$_2$O model \citep[see similar results from][]{Lopez2017}. After several Gyrs of evolution, the radiative-convective boundaries of both planets converge on their interior (metal core $+$ mantle) radii, suggesting an atmosphere that is almost entirely radiative. In this case, and given that both planets have the same surface gravity, the photospheric size of each planet is dependent on their atmospheric mean molecular weight and opacity. The Rosseland mean opacity for H$_2$O is relatively high, particularly in the temperature regime of interest for highly-irradiated super-Earths \citep{Kempton2023}. As such, even a small atmospheric mass fraction $\lesssim 10^{-3}$ can increase a planet's radius by $\sim 10\%$, particularly for small mass planets (see Section \ref{sec:Scenarios}).

\subsection{A Selection of Illustrative End-Member Scenarios} \label{sec:Scenarios}
Our goal is to constrain the water content of super-Earths (see Section \ref{sec:water_content}). To do so, we consider the outcome of two end-member scenarios of planet formation and evolution that can conceivably alter a super-Earth's size for a given mass.  

Firstly, atmospheric escape can alter the size of a super-Earth by removing a significant fraction of its atmosphere. The physics of mass loss for steam-dominated atmospheres and, more generally, high mean-molecular weight atmospheres under high stellar irradiation is currently uncertain. To first order, one would expect a significant fraction of such thin atmospheres to hydrodynamically escape under extreme stellar irradiation. Under such conditions, escaping hydrogen (either molecular, photo-dissociated or ionised) can also drag heavier molecular species \citep[e.g.][]{Hunten1987,Chassefiere1996,Luger2015}. However, the drag efficiency is poorly understood due to its complex dependency on heating sources, e.g. bolometric or X-ray/extreme ultraviolet. Due to these unknowns, we do not explicitly include escape in our models but instead consider the two end-member scenarios, namely the case in which a steam atmosphere is completely retained and the other in which it is completely removed.

Secondly, another process which may alter the size of a super-Earth is outgassing. Sequestered volatiles, such as H$_2$O, reduce the density of interiors since they are much lighter than the background medium, e.g. rock or metal. This is often parameterised as a density deficit. If significantly outgassed, however, the sequestered volatiles are released into the atmosphere. This increases the size of the atmosphere and slightly reduces the size of the interior, due to its increased density. We model the outgassed end-member by forcing the temperature at the top of the mantle to be just below the melting temperature such that the mantle is dry. At the same time, we calculate the amount of water hidden in the core from the equivalent model without outgassing, such that the core contains the maximum amount of water possible. Similar to atmospheric escape, the physics of outgassing on super-Earths is complex. Therefore, we again consider the two end-member scenarios: complete retention of volatiles within a planet's silicate mantle and complete outgassing of volatiles. Note that we do not consider outgassing of volatiles stored in a super-Earth's metal core, since these are very likely to be locked within the core due to inefficient transport from core to planetary surface over Gyr timescales.

As discussed, we choose to model super-Earths with an Earth-like iron mass fraction of $33$\%, due to multiple lines of supporting evidence at the population level \citep[e.g.][]{Adibekyan2021a,Trierweiler2023,CamposEstrada2024}. Nevertheless, changing the iron mass fraction can also alter a planet's bulk density. We discuss this in Section \ref{sec:assumptions}. 

\subsection{Results: Super-Earth Models} \label{sec:ModelResults}

\begin{figure*}
	\includegraphics[width=2.0\columnwidth]{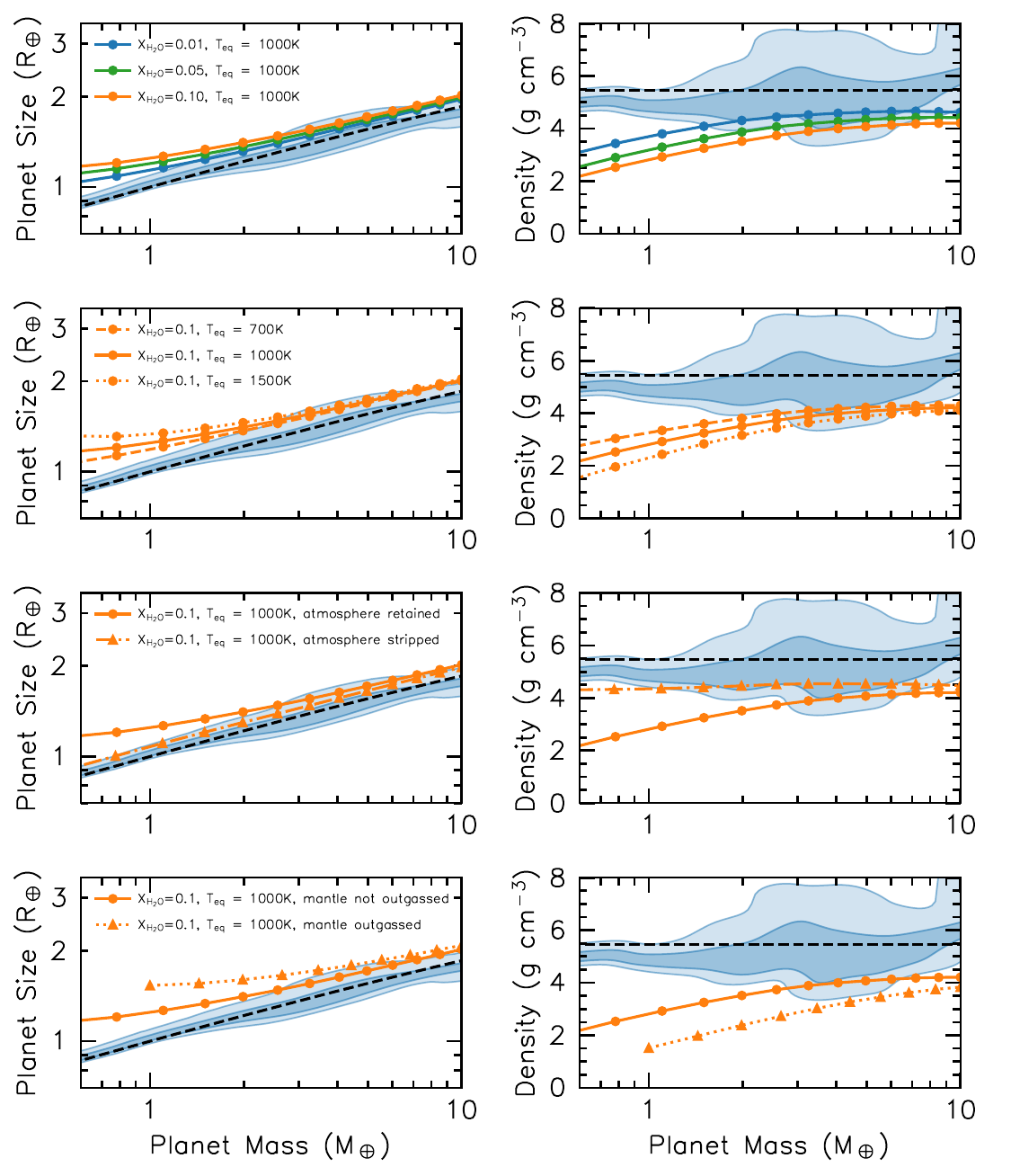}
    \centering
        \cprotect\caption{Super-Earth models are presented alongside the $1\sigma$ and $2\sigma$ contours of observed super-Earths (see Section \ref{sec:DataSelection} and Figure \ref{fig:MRrho}). In the left-hand column, we show this as a function of planet mass versus radius, and in the right-hand column, we show this versus normalised density. In the first row, we show three different total water mass fractions of $0.01$, $0.05$ and $0.10$ in blue, green and orange, respectively. In the second row, we vary the equilibrium temperature from $700$~K, $1000$~K and $1500$~K in dashed, solid and dotted lines, respectively. In the third row, we show models with and without their atmospheres removed as triangles and circles, respectively. In the bottom row, we show models with and without their mantles outgassed in triangles and circles, respectively. Black-dashed lines represent Earth-like with normalised densities $\approx 5.5$ g cm$^{-3}$. {\jr{These mass-radius relations are made publicly available at \href{https://doi.org/10.5281/zenodo.14172976}{10.5281/zenodo.14172976}.}}} \label{fig:Models} 
\end{figure*} 

We present our models in Figure \ref{fig:Models}, all evaluated at $5$~Gyrs and compared to exoplanet observations as described in Section \ref{sec:water_content}.\footnote{{\jr{These models are made publicly available at \href{https://doi.org/10.5281/zenodo.14172976}{10.5281/zenodo.14172976}.}}} In the left-hand panels, we show the masses and radii of our planets under different scenarios. In the right-hand panels, we show the same models as a function of planet mass versus \textit{normalised density}. This is the density a planet would have if it had the same mass as Earth. In other words, planets that are over-dense when compared to Earth sit above $5.5 \text{ g cm}^{-3}$, and vice versa. For this calculation, we utilise an Earth-like grid of models from Section \ref{method:interior} with $33\%$ iron mass fraction and no water content. In the top row, we vary the water mass fraction of the planet from $0.01$ to $0.10$. This water content is partitioned throughout the planet, as described in Section \ref{method:interior}, resulting in steam atmospheric mass fractions varying from $\sim 10^{-5}$ to $\sim 10^{-3}$, with the exact values varying with planet mass.

In the second row, we vary the planet's equilibrium temperature. Atmospheres of a given mass and age at a higher equilibrium temperature are larger than those at lower temperatures. This is because gas densities in the upper radiative atmosphere are reduced, raising the photospheric radius.

In the third row, we demonstrate the effect of atmospheric retention or loss. Dot-dashed triangles represent models for which the steam atmosphere has been completely removed. Note that models with no atmospheres do not contract to Earth-like densities since there is still a significant water content within each planet's interior, resulting in density deficits when compared to Earth.

Finally, in the bottom row, we demonstrate the effect of mantle outgassing. This process pushes more water content into the atmosphere, increasing typical atmospheric mass fractions to several per cent, with the exact value varying with planet mass.

\section{On the Implications of Super-Earth Water Content} \label{sec:water_content}
We now aim to place constraints on the water content of observed super-Earths, utilising the models presented in Section \ref{sec:Method}. We perform this statistical inference under the scenarios presented in Section \ref{sec:Scenarios}, since they are able to change the density of a super-Earth for a given water mass fraction (see Figure \ref{fig:Models}).

\subsection{Sample Selection} \label{sec:DataSelection}

\begin{figure}
	\includegraphics[width=1.0\columnwidth]{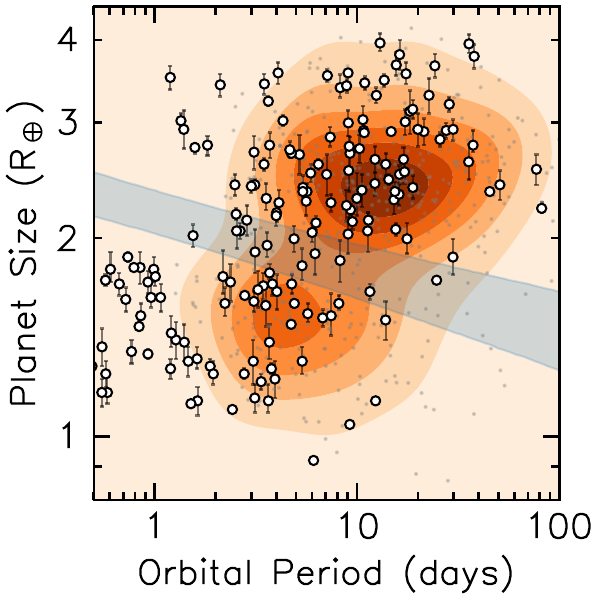}
    \centering
        \cprotect\caption{Our sample of planets from the \verb|PlanetS| catalogue \citep{Otegi2020,Parc2024} are shown as white circles. We distinguish super-Earths from sub-Neptunes using the definition of the radius valley from \citet{Ho2024}, which accounts for the radius dependence of the radius valley on orbital period and stellar mass. Orange contours and grey points show the underlying \textit{Kepler} sample from \citet{Ho2024}. The $1\sigma$ uncertainties on the radius valley are shown in blue, projected through the stellar mass axis.} \label{fig:PerPrad} 
\end{figure} 

\begin{figure*}
	\includegraphics[width=2.0\columnwidth]{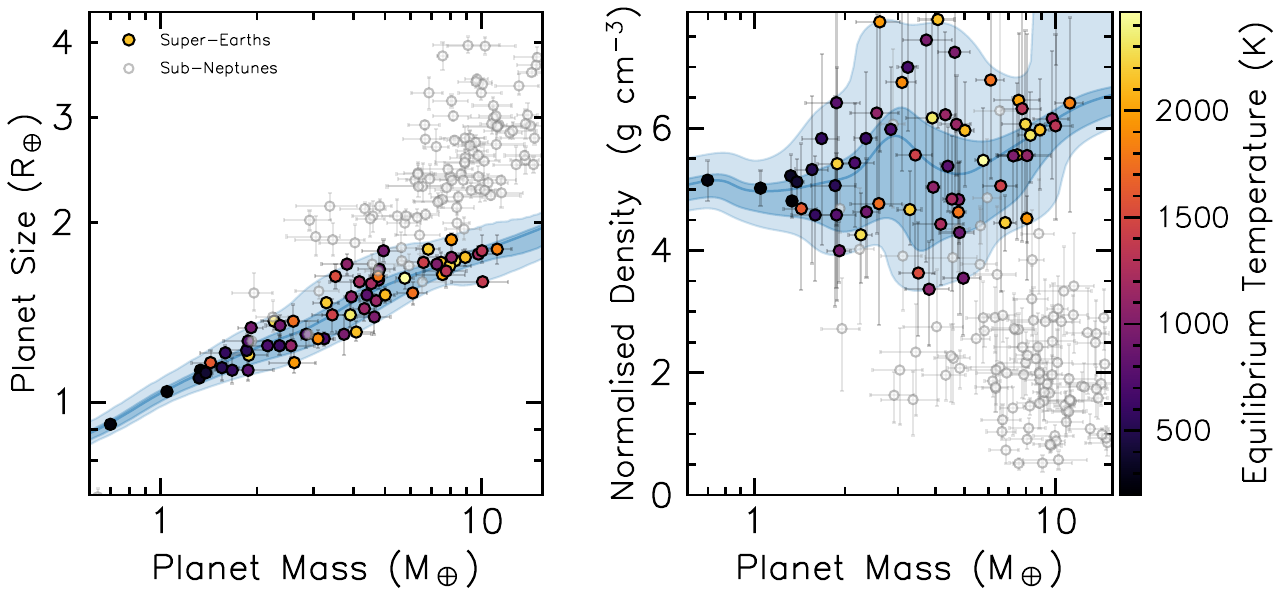}
    \centering
        \cprotect\caption{Our sample of super-Earths is shown as circles, coloured by their equilibrium temperatures. Sub-Neptunes are shown in grey for completeness. Blue contours show the $1\sigma$ and $2\sigma$ bounds on the distribution of super-Earths as a function of planet mass versus radius (left) and normalised density (right). This latter variable represents the density of each planet scaled to Earth, which has a density of $\approx 5.5$ g cm$^{-3}$.} \label{fig:MRrho} 
\end{figure*}

For this analysis, we seek a high-fidelity sample of super-Earths with reliably measured masses and radii. We follow the procedure of \citet{Rogers2024b} in utilising the \verb|PlanetS| catalogue \citep{Otegi2020,Parc2024}, which selects planets based on various reliability and accuracy conditions. Planets in this catalogue have relative mass and radius uncertainties of less than $25\%$ and $8\%$, respectively. Super-Earths are then defined to exist below the radius valley, as prescribed in \cite{Ho2024}. We choose this latter definition of the valley due to its well-defined dependence on orbital period and stellar mass, and its derivation from precise short-cadence \textit{Kepler} data. In Figure \ref{fig:PerPrad}, we present our planet sample radii as a function of orbital period, with the radius valley boundaries projected through the stellar mass axis. We also present the \cite{Ho2024} \textit{Kepler} sample as grey points and orange contours, from which our valley definition is derived. For full details on this data selection criteria, we point the reader to \citet{Rogers2024b}. {\jr{While it is possible to quantify biases for homogenous transit surveys such as \textit{Kepler}, this is typically not possible for the samples including mass measurements such as \verb|PlanetS|. This means that our inferred water mass fractions are representative of the observed population instead of the underlying population.}}

In Figure \ref{fig:MRrho}, we present our sample of super-Earths alongside sub-Neptunes in grey for completeness. In the left-hand panel, we present planet masses and radii, coloured by their equilibrium temperature (assuming zero albedo). In the right-hand panel, we show each planet's normalised density. Blue contours represent the $1\sigma$ and $2\sigma$ boundaries of the distribution of observed planets, calculated with a Kernel Density Estimator (KDE) and equal density in each slice in planet mass.

\subsection{Statistical Methodology} \label{sec:ConstraintsMethod}
Each set of scenario assumptions (i.e. atmospheric stripping, interior outgassing, see Section \ref{sec:Scenarios}) provides a grid of planet models across our adopted parameter space, which spans $M_\text{p} \in [0.5,10] \; M_\oplus$ and $X_{\text{H}_2\text{O}} \in [0.0, 0.4]$, all evaluated at $5$~Gyr. These scenarios are then repeated for 3 different equilibrium temperature regimes $T_\text{eq} \in \{700, 1000, 1500\}$~K. In general, the larger the water mass fraction, $X_{\text{H}_2\text{O}}$, the larger a planet of a given mass due to its inflated interior and more substantial atmosphere (if present). Similarly, the larger the equilibrium temperature, the larger the extent of a steam atmosphere due to decreased gas densities. For each observed super-Earth in our sample, we interpolate over the grid of models to find the water mass fraction reproducing the planet's measured density. In some scenarios, an observed planet may be denser than any of our models, implying an inferred water mass fraction of $X_{\text{H}_2\text{O}}=0$ and possibly a higher iron mass fraction. In other cases, a planet may be less dense than any model, implying a water mass fraction that is greater than any in our grid. In the latter case, which occurs very infrequently, we draw a water mass fraction at random from a uniform distribution between the maximum value in our grid (typically $\sim 0.4$) and $X_{\text{H}_2\text{O}}=1$. This is analogous to a uniform (uninformative) Bayesian posterior beyond the parameter space of our grid of models.

To account for uncertainties in observed planet mass and radius, we iteratively resample with replacement from our population of super-Earths. For each iteration, we repeat the above procedure with observed mass and radii resampled within their $1\sigma$ uncertainties. This sample is then reclassified into super-Earths and sub-Neptunes following Section \ref{sec:DataSelection}, and water mass fractions are calculated as before. We repeat this $10,000$ times for each scenario, yielding $10,000$ measurements of water mass fraction for each super-Earth in our sample. We combine these measurements and quote the $1\sigma$ upper limits on the population-level water mass fraction. We repeat this for model grids at different equilibrium temperatures ($700$~K, $1000$~K, and $1500$~K) so as to investigate the impact of incident flux on inferred water mass fraction. Regardless of the physical scenario, the $1\sigma$ upper limit on water mass fraction equates to a lower limit on normalised density of $4.81$~g cm$^{-3}$ for our sample of super-Earths. 


\begin{figure*}
	\includegraphics[width=2.0\columnwidth]{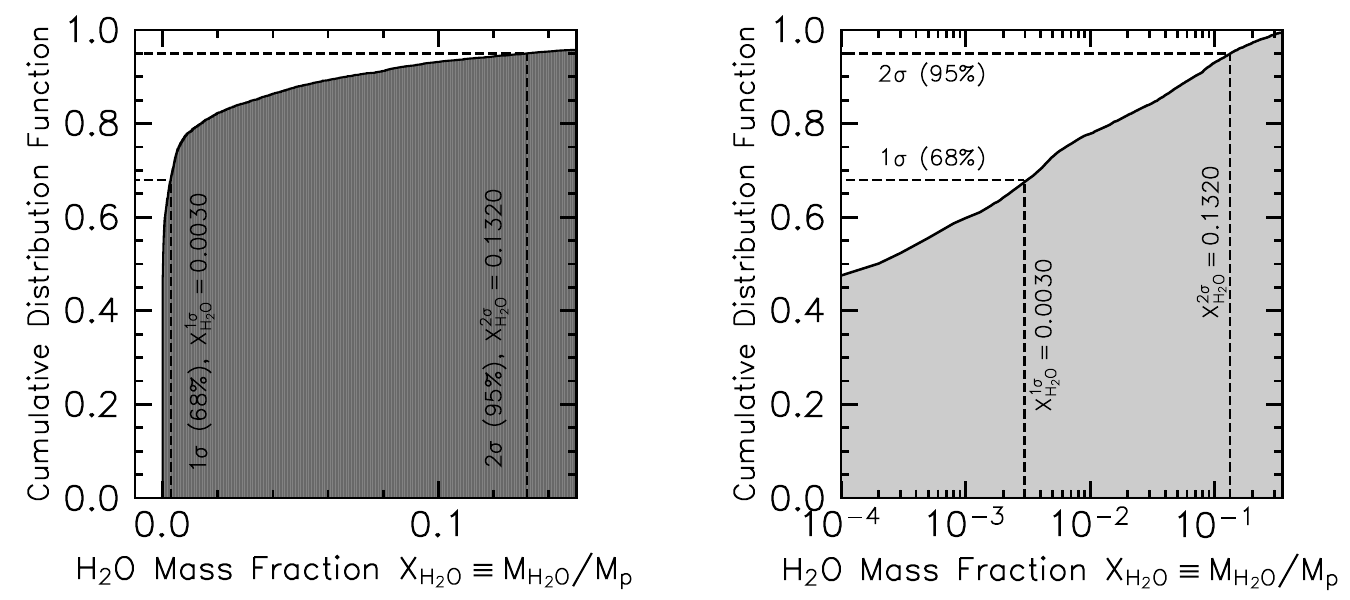}
    \centering
        \cprotect\caption{The distribution of water mass fractions is shown for the population of super-Earths under the scenario of no outgassing and no atmospheric stripping at an equilibrium temperature of $1000$~K. $1\sigma$ and $2\sigma$ limits are also shown.} \label{fig:XH2O_dist} 
\end{figure*} 


\subsection{Results: Super-Earth Water Content} \label{sec:Results_data}
Each modelled scenario yields a distribution of inferred water mass fractions for the population of observed super-Earths. An example can be seen in Figure \ref{fig:XH2O_dist}. Here we show the cumulative distribution function (CDF) of inferred water mass fractions under the scenario of retained steam atmospheres around Earth-like interiors, with no outgassing at an equilibrium temperature of $1000$~K. This CDF is shown with a linear and logarithmic scale on the left and right-hand panel, respectively. One can see that the vast majority of samples are consistent with small water mass fractions $\lesssim 1\%$. We calculate the $1\sigma$ ($68^{\text{th}}$ percentile) upper limit, denoted as $X_{\text{H}_2\text{O}}^{1\sigma} = 0.0030$. For completeness, we also include the $2\sigma$ ($95^{\text{th}}$ percentile), denoted as $X_{\text{H}_2\text{O}}^{2\sigma} = 0.1320$. We highlight that large uncertainties on planet mass and radius are likely driving the $2\sigma$ value. Hence, when quoting upper limits of water mass fraction, we choose the $1\sigma$ value. 

\begin{table*}
\begin{center}
\begin{tabular}{ |c|c||c|c| } 
 \hline
  & $T_\text{eq}$ (K) & No Outgassing & Complete Outgassing \\
  \hline\hline
 \multirow{3}{*}{Atmosphere Retained} & $700$~K & $<0.0048$ & $<0.0037$ \\
 & $1000$~K & $<0.0030$ & $<0.0023$ \\
 & $1500$~K & $<0.0024$ & $<0.0012$ \\
 \hline
 Atmosphere Stripped & - &  $<0.0333$ & - \\ 
 \hline
\end{tabular}
\end{center}
\caption{Upper limits ($1\sigma$) on water mass fraction for observed super-Earths. We consider various scenarios for planets with Earth-like iron mass fractions with or without undergoing atmospheric escape or mantle outgassing. We also consider models at various equilibrium temperatures of $700$~K, $1000$~K and $1500$~K. Note that constraints are independent of equilibrium temperature if the planet's atmosphere has been stripped. No constraints can be placed in the scenario of complete mantle outgassing \textit{and} complete atmospheric stripping.} \label{tab:Limits} 
\end{table*}

\begin{figure}
	\includegraphics[width=1.0\columnwidth]{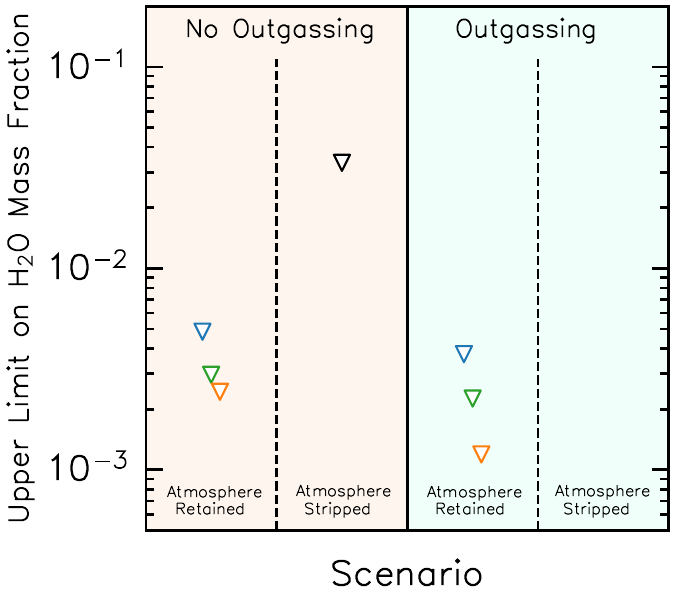}
    \centering
        \cprotect\caption{Summary of population-level water mass fraction $1\sigma$ upper limits for the observed sample of super-Earth. Scenarios are laid out as follows (from left to right): no outgassing and no atmospheric stripping; no outgassing with atmospheric stripping; outgassing without atmospheric stripping; and finally outgassing and atmospheric stripping. For scenarios that exhibit a dependence on the assumed equilibrium temperature of the sample, constraints at $700$~K, $1000$~K and $1500$~K are shown in blue, green and orange, respectively.} \label{fig:Summary} 
\end{figure} 

We present a summary of water mass fraction upper limits for various scenarios with Earth-like iron mass fractions in Figure \ref{fig:Summary}, as well as in Table \ref{tab:Limits}. 

\vspace{1cm}
\subsubsection{Atmospheres Retained, No Outgassing}
Starting with the scenario of retained atmospheres and no mantle outgassing, we find an upper limit of $\sim 0.3\%$, with a weak dependence on equilibrium temperature. This is due to the fact that planet atmospheres of a given mass and age are more inflated at higher equilibrium temperatures. Our upper limits on water mass fraction are thus lower for higher equilibrium temperatures, which yield the same observed density for lower water content. For a typical \textit{total} water mass fraction of $\sim 0.3\%$, the \textit{atmospheric} mass fraction can be as low as $\sim 10^{-6}$, or equivalently $\sim 0.1$~bar at its base. Recall from Section \ref{method:atmosphere} that water has a large Rosseland mean opacity, meaning that only a small atmospheric mass is required to increase a planet's transit radius significantly. This effect is more noticeable for smaller planet masses (see Figure \ref{fig:Models}).

\begin{figure*}
	\includegraphics[width=2.0\columnwidth]{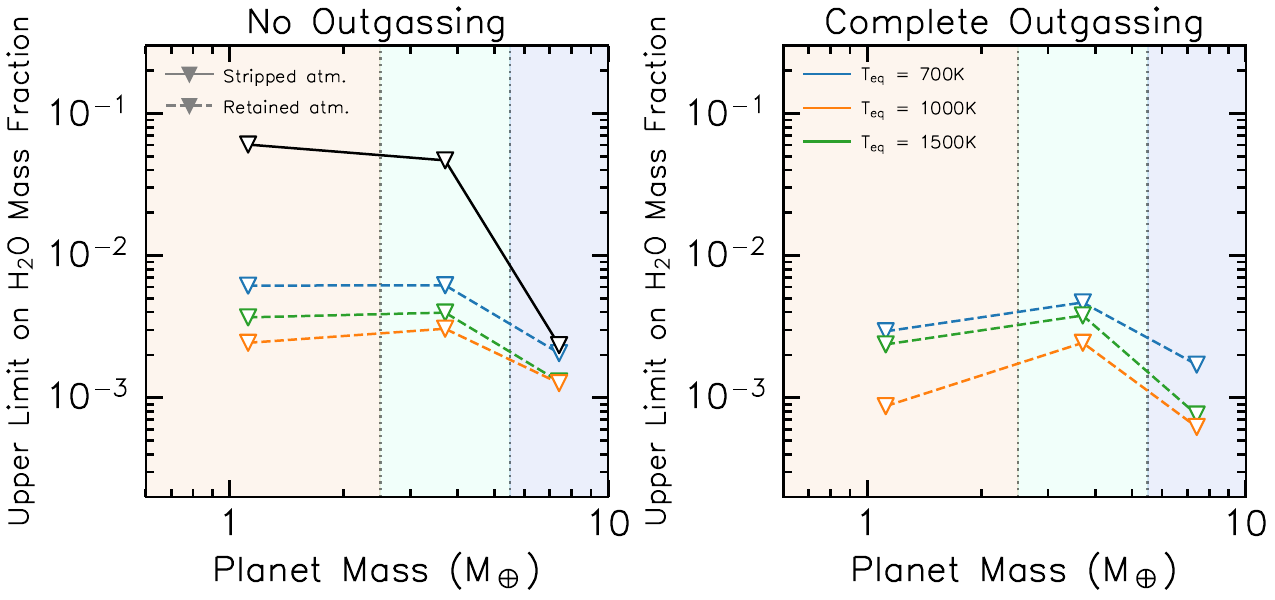}
    \centering
        \cprotect\caption{Water mass fraction upper limits are shown as a function of planet mass. Shaded regions represent the bin widths in planet mass, with edges at $M_\text{p} = [0.5, 2.5, 5.5, 10.0] \, M_\oplus$. Scenarios with and without mantle outgassing are shown in the right and left-hand panels, respectively. Scenarios without atmospheric stripping exhibit a dependence on the assumed equilibrium temperature, which we show for $700$~K, $1000$~K and $1500$~K in blue, green and orange, respectively.} \label{fig:MassBins} 
\end{figure*} 

\subsubsection{Atmospheres Retained, Mantles Outgassed}
Now, with the inclusion of mantle outgassing in this scenario, we see that our upper limits are reduced further to $\lesssim 0.2\%$. This is because releasing the water content of a planet's mantle into its atmosphere increases its size. As a result, an even lower water mass fraction is required to match the observed densities of super-Earths. As with the previous scenario, the same dependence with equilibrium temperature exists.

\subsubsection{Atmospheres Stripped, No Outgassing}
The scenario of atmospheric stripping without mantle outgassing provides an upper limit on water mass fraction of $\lesssim 3\%$. This scenario is the most stringent since there is no remaining atmosphere to inflate the planet. Instead, a planet's density is controlled by varying the water content locked inside its mantle and iron core. Since the compressibility of a gas is far greater than that of silicate/iron, more water is required to achieve the same reduced density. Note that water mass fractions for scenarios with atmospheric stripping are independent of equilibrium temperature since there is no atmosphere to inflate/contract as a function of incident flux. 

\subsubsection{Atmospheres Stripped, Mantles Outgassed}
Our final scenario, which comprises total mantle outgassing \textit{and} total atmospheric stripping, yields no constraints. This is because, from an initial maximum water mass fraction in our scenario grids of $X_{\text{H}_2\text{O}} \sim 0.4$, outgassing and stripping lead to the water content being reduced to $X_{\text{H}_2\text{O}} \lesssim 0.05$, which is completely stored in the planet's iron core. As mentioned previously, the relative lack of compressibility in the iron core means that water mass fractions of $\gtrsim 0.4$ (before stripping) are required to fit the observed densities of super-Earths. Since our models are unable to reach such values in a reliable manner, we cannot place limits on this scenario.

\subsection{Mass Dependent Limits of Water Mass Fraction}
We now split our sample of super-Earths into three bins in planet mass and find the upper limits of water mass fraction for each bin. We choose bin edges of $M_\text{p} = [0.5, 2.5, 5.5, 10.0] \, M_\oplus$, such that an approximately equal number of super-Earths sit within each bin. We present results in Figure \ref{fig:MassBins}, with bins shaded in different colours for clarity. The right and left-hand panels show scenarios with and without outgassing, respectively.

We see the same trends as in Figure \ref{fig:Summary}, with scenarios involving atmospheric stripping providing systematically larger upper limits. Broadly speaking, we see that planets $\lesssim 5.5 M_\oplus$ have larger upper limits on water content than those of larger masses, especially under the scenario of atmospheric stripping without outgassing. As discussed, a greater dynamic range in water mass fraction is needed to account for varying planet density under this scenario, since there is no atmosphere to inflate the planet. When considering the entire population, the constraint under this scenario is $X_{\text{H}_2\text{O}} \lesssim 0.2\%$, however, when divided into mass bins, we find an upper limit of $\sim 5\%$ in the lowest mass bin, approximately $25\times$ higher than in the highest mass bin.

\section{Discussion} \label{sec:Discussion}

Across our various end-member scenarios for super-Earths, including processes such as atmospheric escape and outgassing, we generally find upper limits on water mass fractions $\lesssim 3\%$, as shown in Figure \ref{fig:Summary}. We now discuss the implications of these findings on their formation and evolution.

\subsection{Distinguishing between scenarios} 
Although end-member scenarios are informative, the true upper limit on water mass fraction cannot be known without an understanding of which of our scenarios best describes reality. 


Perhaps the best way to distinguish between scenarios with and without atmospheric stripping is with transit and emission spectroscopy observations. The upcoming 500 hours of Cycle 3 Director Discretionary Time (DDT) with \textit{JWST} \citep{Redfield2024} aims to determine if rocky planets around M-dwarfs host atmospheres. If atmospheres are indeed detected around planets close to their stars, then it implies mass loss of primary or secondary atmospheres is not completely efficient, perhaps due to efficient line cooling of molecular gas species, inefficient drag from lighter species, or outflows becoming non-collisional. If, on the other hand, evidence is found for the lack of atmospheres,\footnote{A corollary of this finding would be that mass loss timescales are shorter than outgassing timescales.} then it implies efficient escape and that more water is perhaps stored inside super-Earths, as shown in Figures \ref{fig:Summary} and \ref{fig:MassBins}. Although we have only focused on steam atmospheres in this study, we highlight that it is the mean molecular weight and opacities that largely control the extent of evolved atmospheres (see Figures \ref{fig:H2_vs_H2O_evolution} and \ref{fig:Models}). Inference of other high mean-molecular weight species (e.g. CO, CO$_2$, CH$_4$, NH$_3$) will also provide crucial information that can be used to inform super-Earth water budgets in future studies.


Mantle outgassing is arguably the more difficult scenario to place constraints on since it is poorly understood beyond the context of solar-system bodies and depends on many factors, such as the degree of magma ocean crystallisation and the presence of crust \citep[e.g.][]{Crisp1984,Dorn2018b,Gaillard2021,Guimond2021,Krissansen-Totton2022}. Typically, outgassing is investigated in the context of Earth-like carbon cycles; however, for super-Earths with larger water mass fractions and potentially deep ocean layers, the abundance of H$_2$O may change the oxidation state of interiors, thus changing chemical signatures of outgassing. Again, atmospheric characterisation may shed light on this topic, although we highlight that more theoretical work is needed to interpret such observations.

We speculate that the most realistic end-member scenario, given our current understanding, is that of atmospheric stripping without mantle outgassing. Whilst outgassing is likely an important process in terrestrial planet evolution, mantles are unlikely to \textit{completely} outgas. Regarding atmospheric stripping, super-Earths typically exhibit extreme equilibrium temperatures and are unlikely to retain a significant, optically-thick, steam-dominated atmosphere. H$_2$O is likely to photo-dissociate, allowing hydrogen to drag oxygen along in the outflow. Of course, these scenarios are end-members, and reality will likely be a mixture of processes, e.g. partial outgassing and near-complete stripping. Nevertheless, the combination of our scenario results suggests upper limits of order $\sim 1\%$. In order for a larger water mass fraction to have been present, \citep[such as those in proposed sub-Neptune ``water-worlds'' with typical fractions of $\gtrsim 50\%$ e.g.][]{Zeng2019,Luque2022,Burn2024}, one would need near-total outgassing of mantle water and total atmospheric stripping. We encourage further studies to investigate the likelihood of such a scenario.

\subsection{On the origins of water} \label{sec:WaterOrigins}

\begin{figure}
	\includegraphics[width=1.0\columnwidth]{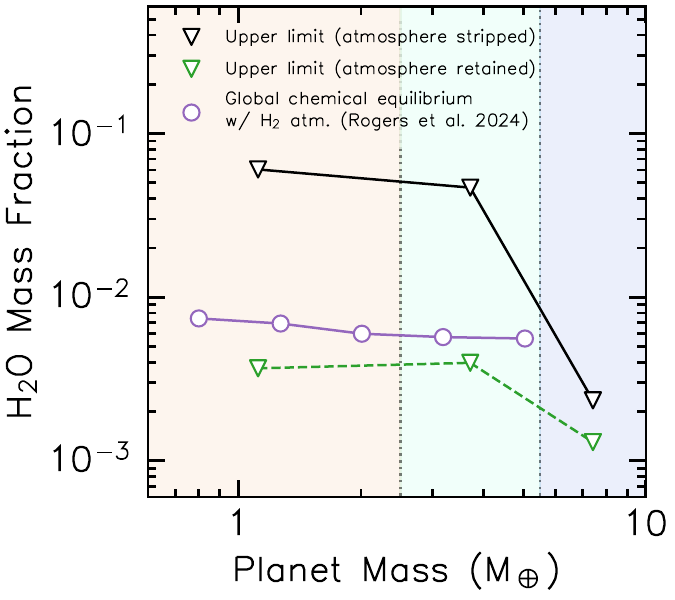}
    \centering
        \cprotect\caption{Similar to the left-hand panel of Figure \ref{fig:MassBins}, but compared to the models of \citet{Rogers2024b}. These models determine the water mass fraction produced as a result of the chemical equilibration of primordial hydrogen-dominated atmospheres with magma oceans.} \label{fig:Comp_wRogers2024} 
\end{figure} 

Throughout this study, we have been agnostic as to the origin of water on super-Earths. One mechanism capable of providing an abundance of water is via delivery through pebble or planetesimal accretion \citep[e.g.][]{Machida2010,Sato2016,Ida2019,Steinmeyer2023}. However, another route to water is through its production as a result of chemical reactions. In \citet{Rogers2024b}, super-Earths were modelled under the assumption that they began life as sub-Neptunes, with primordial hydrogen-rich atmospheres accreted from their nascent protoplanetary discs. Using the models of \citet{Schlichting2022,Young2023}, the authors solved for global chemical equilibrium on super-Earths, accounting for the loss of hydrogen as it was removed via thermally driven atmospheric escape. They found that a large fraction of water can be produced as a result of hydrogen chemically reacting with a magma ocean, which is then primarily stored in the planet's iron core as atomic H and O. Figure \ref{fig:Comp_wRogers2024} compares these models, which were performed for super-Earths at equilibrium temperatures of $1000$~K, with our upper limits under the scenarios of no outgassing at the same equilibrium temperature. In general, we see good agreement with the models of \citet{Rogers2024b}, particularly in the case of atmospheres being stripped. This suggests that the water content of most super-Earths can be explained as the result of chemical equilibration with a primordial hydrogen-dominated atmosphere. This is similar to the argument of \citet{Young2023}, who argued that Earth's water content could be explained as a result of a primordial hydrogen-dominated atmosphere equilibrating with a magma ocean \citep[see also][]{Wu2018}. In the sub-Neptune regime, \citet{Benneke2024} showed that the mean-molecular weight of TOI-270b's atmosphere, as observed by \textit{JWST}, can be explained by the production of H$_2$O (among other volatile species) via the equilibration of its hydrogen dominated atmosphere with a magma ocean.

Placing super-Earths in the context of all known small planets, including Earth, and combining with the observational evidence in favour of hydrogen-dominated atmospheres on small exoplanets \citep[e.g.][]{JontofHutter2014,DosSantos2023a} potentially leads to a concise story for the origin of water via chemical reactions with a magma ocean. This is backed up by numerous experimental studies, highlighting the efficient production of water when hydrogen reacts with silicate magma \citep[e.g.][]{Shinozaki2014,Kim2023,Horn2023b,Horn2023a}.


\subsection{Benchmarking the models} \label{sec:benchmarking}
In Section \ref{sec:WaterOrigins}, we compared our models with those of \citet{Rogers2024b}. Here, we highlight that there are notable distinctions in the interior models adopted in this work from \citet{Luo2024} (see Section \ref{method:interior}) from those of \citet{Rogers2024b}. The latter solves a complex network of chemical equilibria throughout the planet; however, it is limited in the complexity regarding interior structure (e.g. pressure and temperature profiles). The models of \citet{Luo2024}, on the other hand, solve for pressure and temperature profiles but only track water as a chemical component. Furthermore, both models adopt different H$_2$O solubility law scalings, with those of \citet{Luo2024} assuming H$_2$O dissolves in a magma ocean as OH$^-$, as suggested by experiments at low pressure \citep[e.g.][]{Hamilton1964}. \citet{Schlichting2022,Rogers2024b}, on the other hand, assumed ideal mixing of H$_2$O in a magma ocean due to the lack of evidence for the former being true at high pressures. Finally, \citet{Luo2024} assumed partitioning coefficients of H$_2$O between mantle and iron core that assume water remains in its molecular form, whereas \citet{Schlichting2022,Rogers2024b} assume H$_2$O speciates into atomic H and O due to the extreme temperatures and pressures. 

Due to these differences, we performed a simple comparison to confirm that both interior models yield similar inferences of total water mass fraction for the observed population of super-Earths. With the goal of consistency, we adapted the models of \citet{Schlichting2022,Rogers2024b} to allow for water dissolving in silicate melt as OH as in \citet{Dorn2018,Luo2024}. We then produced a small grid of planet models of varying water mass fraction at $4$~M$_\oplus$ at an equilibrium temperature of $1000$~K and evolve their steam-dominated atmospheres with the method described in Section \ref{method:atmosphere}. We then repeated the water mass fraction upper limit calculation (as described in Section \ref{sec:ConstraintsMethod}) for a small sample of observed super-Earths $3 \leq M_\text{p} / M_\oplus < 5$. We found that the upper limit with the interior models of \citet{Rogers2024b} is $\sim 0.9\%$, compared to $0.5\%$ for models constructed in this work. This small difference is reassuring, given the differences in formalism between the models of \citet{Rogers2024b} and \citet{Luo2024}. 

\subsection{Assumptions and limitations} \label{sec:assumptions}
The results of this study have relied upon interior models of rocky super-Earths. These models are built upon experimentally, or computationally derived partitioning coefficients, solubility scalings and equations of state. It is clear that more work is needed to refine and validate our understanding of volatile sequestration in rocky planets. Furthermore, our models have been limited to a single volatile species, namely H$_2$O. In reality, the chemistry involved is far more complex, with other oxygen, carbon, nitrogen and sulphur-bearing species likely present in atmospheres and interiors. Studying their behaviours in atmospheres, as well as under high pressures and temperatures, is crucial in order to move beyond simple models.

{\jr{An important simplifying assumption of this work has been to assume Earth-like iron mass fractions for super-Earths. Whilst this iron-to-silicate ratio is supported by various lines of evidence, as discussed in Sections \ref{sec:intro} and \ref{sec:Scenarios}, it is worth noting that changing the iron mass fraction of a planet also changes its bulk density. Adopting the same approach as with our other scenarios, we consider the end-members of this change. We performed our statistical analysis again under the assumption that super-Earths have Mercury-like iron mass fractions of $69\%$. The increased iron mass fraction increases the density of a planet's interior for a given mass. Thus, larger water mass fractions are required to match observed super-Earth densities. In the case of no atmospheric stripping and no outgassing, we find $X_{\text{H}_2\text{O}}^{1\sigma} < 0.1723$, $0.1477$ and $0.1168$ for equilibrium temperatures of $700$~K, $1000$~K and $1500$~K, respectively. In the case of no stripping and complete mantle outgassing, we find $X_{\text{H}_2\text{O}}^{1\sigma} < 0.0813$, $0.0569$ and $0.0353$ for equilibrium temperatures of $700$~K, $1000$~K and $1500$~K, respectively. As expected, these are larger than those of Earth-like iron mass fractions (see Table \ref{tab:Limits}), although the physical validity of this scenario is somewhat unlikely, since it would likely require mantle stripping via dynamically induced collisions \citep[e.g.][]{Benz1988}, which would then remove any atmosphere. There is also a lack of evidence for the existence of super-Mercuries in polluted white dwarf data \citep{Doyle2020}. We were unable to find reasonable constraints on water mass fractions in the case of atmospheric stripping since an extremely large water mass fraction is required to match the densities of observed super-Earths. The opposite end-member scenario consists of planets without iron cores. We find that this scenario is inconsistent with the vast majority of super-Earths, which are denser than a pure (dry) silicate composition and thus require a water mass fraction of $X_{\text{H}_2\text{O}} = 0$. The population-level upper limits would thus be very small $X_{\text{H}_2\text{O}}^{1\sigma} < 10^{-4}$ and uninformative.}}

Finally, our inferred water mass fraction limits have also relied up on the precise characterisation of observed super-Earths. We highlight, however, that this sample is inherently biased since radial velocity programs preferentially target stars with larger signals. Our constraints are, therefore, indicative of the \textit{observed} population and not the \textit{underlying} population. We encourage future studies to address survey bias to quantify this effect.  In addition, increasing the precision of mass measurements, and increasing the available sample size with missions such as \textit{TESS} \citep{Ricker2015} and \textit{PLATO} \citep{Rauer2014} will refine our inferences in future studies.

\section{Conclusions} \label{sec:Conclusions}
In this work, we have focused on the water content of super-Earths. We combine the interior models of \citet{Luo2024} with atmospheric evolution models to construct grids under various end-member evolutionary outcomes, including efficient mantle outgassing and/or atmospheric escape. {\jr{These mass-radius relations are made publicly available at \href{https://doi.org/10.5281/zenodo.14172976}{10.5281/zenodo.14172976}.}} By comparing these models with an observed sample of super-Earths from the \verb|PlanetS| catalogue \citep{otegi_impact_2020,Parc2024}, we place upper limits on the water mass fractions of super-Earths at the population level. Our conclusions are as follows:

\begin{itemize}
    \item In the scenario that super-Earths retain a thin steam atmosphere, we find a $1\sigma$ upper limit on water mass fraction to be $\sim 0.3\%$, with a weak dependence on the assumed equilibrium temperature.
    \item In the scenario that super-Earths do not host atmospheres, possibly due to thermally driven escape, this water mass fraction upper limit increases to $\sim 3\%$. 
    \item In the scenario that super-Earths outgas the entire water content of their mantles into a steam atmosphere, the upper limit reduces further to $\sim 0.2\%$. If this outgassed steam atmosphere is stripped, then we are unable to place any constraint on the original water mass fraction. 
    \item Larger super-Earths $\gtrsim 5 M_\oplus$ tend to have systematically lower water mass fractions than those with lower masses. 
\end{itemize}

Our study is designed to consider end-member scenarios, with reality likely sitting somewhere in between, e.g. partial outgassing and significant atmospheric escape. Nevertheless, our combined water mass fraction upper limits are consistent with studies suggesting that reactions between primordial hydrogen-dominated atmospheres and magma oceans can produce $\sim 1\%$ water \citep{Rogers2024b}. Similar arguments have been proposed for Earth's water budget \citep{Young2023}, suggesting that the water content of small planets may be produced, instead of delivered.


\section*{Acknowledgements}
{\jr{We kindly Rafael Luque, Joseph Murphy, and the anonymous reviewer for comments that helped improve the paper.}}. JGR is sponsored by the National Aeronautics and Space Administration (NASA) through a contract with Oak Ridge Associated Universities (ORAU). HES gratefully acknowledges support from NASA under grant number 80NSSC21K0392 issued through the Exoplanet Research Program. For the purpose of open access, the authors have applied a Creative Commons Attribution (CC-BY) licence to any Author Accepted Manuscript version arising. C.D. acknowledges support from the Swiss National Science Foundation under grant TMSGI2\_211313. E.D.Y. acknowledges support from the AEThER program funded by the Alfred P. Sloan Foundation grant G202114194.  This work has been carried out in parts within the framework of the NCCR \verb|PlanetS| supported by the Swiss National Science Foundation under grant 51NF40\_205606.

\bibliography{references,referencesCaro}{}
\bibliographystyle{aasjournal}




\end{document}